# Three-stage dynamics of nonlinear pulse amplification in ultrafast mid-infrared fiber amplifier with anomalous dispersion

Weiyi Sun,[a,b,c] Jiapeng Huang,[a,*] Liming Chen,[a] Zhuozhao Luo,[a] Wei Lin,[a,d] Zeqing Li,[a,b] Cong Jiang,[a] Zhiyuan Huang,[b] Xin Jiang,[a] Pengfei Wang,[e,*] Yuxin Leng,[b] and Meng Pang,[a,b,d,*]

a. Russell Centre for Advanced Lightwave Science, Hangzhou Institute of Optics and Fine Mechanics and Shanghai Institute of Optics and Fine Mechanics, Hangzhou, 311421, China
b. State Key Laboratory of High Field Laser Physics and CAS Center for Excellence in UltraIntense Laser Science, Shanghai Institute of Optics and Fine Mechanics (SIOM), Chinese Academy of Sciences (CAS), Shanghai 201800, China
c. School of Physical Science and Technology, Shanghai Tech University, Shanghai 200031, China
d. School of Physics and Optoelectronic Engineering, Hangzhou Institute for Advanced Study, University of Chinese Academy of Sciences, Hangzhou 310024, China.
e. Center for Advanced Optoelectronic Functional Material Research, Northeast Normal University, Changchun, 130024, China

**Abstract:** Nonlinear pulse amplification in optical fiber, with capability of breaking the gain-bandwidth limitation, is a key technique for high-energy, ultrafast pulse generation. In the longer wavelength region (including 1.55 μm, 2 μm and 2.8 μm) where the gain fiber has normally strong anomalous dispersion, the nonlinear amplification process over fiber exhibits more complicated dynamics than that of its 1-μm counterpart, and the underlying mechanism of the nonlinear pulse propagation process in high-gain anomalous fiber is still elusive so far. Here, we demonstrate an in-depth study on the nonlinear amplification process in high-gain ultrafast mid-infrared fiber, providing clear physical understanding on the debate of adiabatic soliton compression. We unveil that under the high-gain condition, the ultrafast pulse launched into the anomalous gain fiber experiences successively three distinct stages, named as the balance between linear and nonlinear chirp, high-order-soliton-like pulse compression and pulse splitting due to high-order effects. While a relatively-clean ultrafast pulse can be obtained immediately after the high-order-soliton-like compression stage, excessive gain fiber length could hardly enhance further the pulse peak power due to soliton splitting. Our findings can provide several critical guidelines for designing high-power ultrafast fiber amplifiers at near- and mid-infrared wavelengths.

**Keywords**: middle infrared fiber laser，nonlinear fiber optics，ultrafast optics，mode-locked laser，laser amplifier

*Address correspondence to Jiapeng Huang, E-mail: jiapenghuang@siom.ac.cn; Pengfei Wang, E-mail: pengfei.wang@tudublin.ie; Meng Pang, E-mail: pangmeng@siom.ac.cn

## 1 Introduction

Nonlinear amplification of ultrafast pulses in gain fiber is an important technique for high-energy pulse generation with ultrashort pulse width. [1-4] Two unique advantages of this technique, compared with the scheme of conventional chirped pulse amplification (CPA), [5] render it a large amount of research attention in fields of ultrafast lasers and nonlinear fiber optics. [6,7] The first one is that through properly utilizing, rather than suppressing, nonlinear effects in gain fiber, optical spectrum of the amplified pulse output from the fiber could be much broader than that of the input pulse, and even broader than the gain bandwidth, giving rise to the possibility of shorter pulse generation. [6,8,9] The typical example is the well-known parabolic pulse amplification scheme, [1,2,9] in which elegant collaborations among optical gain, normal fiber dispersion and Kerr nonlinearity lead to the self-similar evolution of the amplified pulse, resulting in broadband output spectrum with almost-linear chirp that can be perfectly compressed using a grating-based compressor. [10-12] The second advantage is that in some nonlinear amplification schemes based on chirp pre-compensation or dispersion-engineering techniques,[12-14] high-energy chirp-free pulses can be directly obtained at the output port of the gain fiber, avoiding the use of pulse compressor. Such schemes largely simplify the configuration of high-energy ultrafast pulse amplifier, paving the way to monolithic femtosecond fiber laser systems. [13,15]

Most of nonlinear amplification systems at ~1 μm wavelengths rely on relatively-explicit operation principle, since the fiber material (silica glass) exhibits normal dispersion in this wavelength regime. In normal-dispersion gain fiber, the high-frequency component of the pulse has a lower group velocity than the low-frequency one, moving toward the trailing edge of the pulse. The index modulation at the pulse trailing, due to Kerr nonlinearity, leads to a blue shift of the high-frequency component toward the higher-frequency side. [7,16] Therefore, the collaboration between normal dispersion and self-phase modulation results in a continuous broadening of both the pulse spectrum and its temporal width. [7,13] In contrast, as the operation wavelength of the amplifier moves to longer wavelengths, for example to ~1.55 μm, [17,18] ~2 μm [15,19] or ~3 μm, [14,20,21] the physical picture of the nonlinear amplification changes dramatically, since the glass material has anomalous dispersion at these wavelengths. [22] In a few previous experiments, temporally-compressed ultrafast pulses have been obtained directly at the output port of the gain fiber with no need of additional chirp compensation.[23,24] However, the underlying mechanism of such a nonlinear amplification process under anomalous dispersion landscape is still inexplicit so far, and Raman self-frequency shifts and non-trivial pulse pedestals were frequently observed in such nonlinear amplification systems,[10,25] seriously limiting the quality of the output pulses from the amplifiers.

Even though the concept of soliton amplification and compression,[13] based on the adiabatic soliton propagation,[26] were widely discussed for understanding the nonlinear amplification process in strong anomalous gain fiber, the critical debate still lies in the relative scales between the gain length ($L_g$) and the nonlinear length ($L_n$) in such a system, as expressed in Equation 1, where $g$ is the gain coefficient, $\gamma$ the Kerr nonlinearity coefficient and $P_0$ the pulse peak power. While the adiabatic soliton theory relies on the basic assumption of $L_g \gg L_n$, highly-doped gain fiber samples (with high gain coefficients) were usually used in practical experiments, [10,13] leading to short gain lengths of <1 m. Therefore, the adiabatic criterion is normally hard to be satisfied in these experiments.[25, 27]

$$L_g = \frac{1}{g} \qquad L_n = \frac{1}{\gamma P_0} \tag{1}$$

The large deviation from the adiabatic criterion would lead to, in one side, more complicated dynamics of nonlinear amplification process.[25] In the other side, some guidelines for the amplifier design, based on the adiabatic assumption, may become invalid. For example, in order to ensure the adiabatic propagation a longer gain fiber length should be applied for obtaining a higher energy pulse. Moreover, according to the adiabatic soliton theory, the pre-engineering technique of pulse chirp could hardly improve the system performance, since in principle a Fourier-transform-limited soliton without chirp is more ideal as a seed pulse to enter the trajectory of adiabatic soliton propagation.

In this article, we demonstrate that at the high-gain (short-gain-length) condition, the nonlinear amplification process of ultrafast pulse in anomalous dispersion fiber can be divided into three distinct stages. In the first stage, the linear pulse chirp accumulated at the beginning of the propagation could be compensated gradually by the nonlinear chirp due to self-phase modulation, as the pulse energy increases in the gain fiber. The output pulse from the first stage has nearly zero chirp, and can be regarded as a high-order soliton launched into the second stage. In the second stage, the process like high-order-soliton compression [28] can be obtained, in which the pulse spectrum is largely broadened with its temporal width significantly shortened over a quite short length of gain fiber. Then, the increased pulse peak power and enlarged spectral bandwidth dramatically enhance the effects of both higher-order ($\geq 3$) dispersion and high-order nonlinear effects in the gain fiber, breaking the system symmetry. In the third stage, soliton splitting [13] is therefore observed, leading to large pulse pedestal and obvious Raman-induced soliton self-frequency shift. We unveil both experimentally and numerically the nonlinear amplification dynamics in an ultrafast ZBLAN-fiber amplifier operating at 2.8 μm, and demonstrated the universality of the three-stage model. Our results clarify some limitations of adiabatic soliton amplification theory, and provide a few useful guidelines for constructing near- and mid-infrared ultrafast fiber amplifiers with improved pulse qualities.

## 2 Experimental set-up and results

The experimental set-up we constructed is illustrated in Fig. 1(a). The seed oscillator is a conventional mid-infrared soliton fiber laser mode-locked through nonlinear polarization rotation (NPR).[29,30] When pumped with a high-power laser diode at 975 nm, the highly-doped $Er^{3+}$:ZBLAN double-cladding gain fiber provides optical gain at ~2.8 μm. The gain fiber length is 6.5 m, and the cavity round-trip frequency is ~30 MHz. Without any dispersion-management element in the laser cavity, the strong anomalous dispersion ($\beta_2$ = -86 ps$^2$/km) of the gain fiber forces the laser to operate in the soliton regime.[31] Through adjusting the waveplates inside the laser cavity, stable mode-locking pulse train can be obtained at a pump power of ~5 W, leading to the generation of soliton-like pulse at the output port (30% dichroic mirror) of the laser, see Fig. 1(a). At this condition, we measured the laser output power to be ~33 mW, corresponding to a pulse energy of ~1.1 nJ. In the experiment, the laser spectrum was measured using a Fourier transform infrared spectrometer (FTIR), and the results are illustrated in Fig. 1(b), giving a 3-dB bandwidth of 8.7 nm. The presence of strong Kelly sidebands on both sides of the laser spectrum indicates that the laser is operating in the soliton regime.[32] The pulse duration was measured using a second harmonic generation (SHG) autocorrelator with a temporal resolution of 10 fs, and the results are illustrated in Fig. 1(c). The temporal width of the pulse was measured to be 998 fs, and therefore the time-bandwidth product (TBP) was calculated to be 0.336, further verifying the soliton operation of the seed laser.[28]

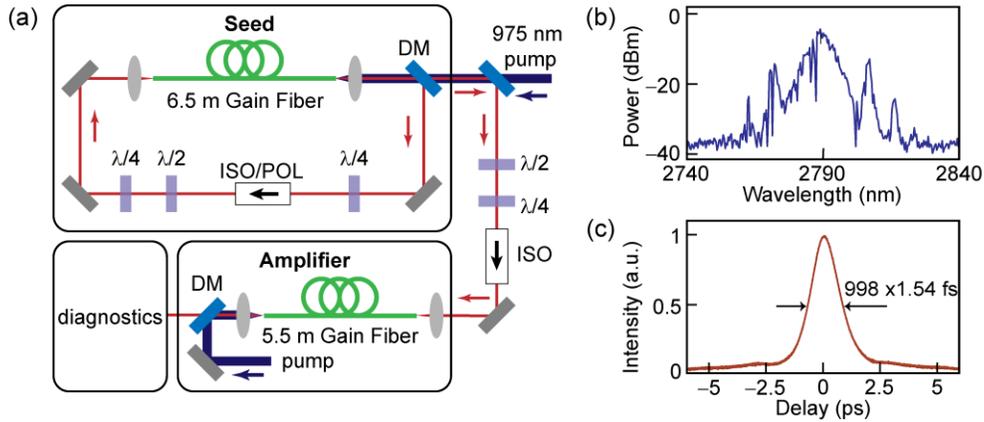

**Fig. 1** (a) Er-doped ZBLAN fiber laser system (see text for details). ISO, isolator; DM, dichroic mirror; λ/2, half-wave plate; λ/4, quarter-wave plate. (b) and (c) Measured optical spectrum and autocorrelation trace of the seed pulse.

The output pulse from the seed laser passed through two waveplates for controlling the light polarization state and then an optical isolator for preventing the back-reflection, and the output pulse from the isolator with an energy of ~0.7 nJ was launched into a section of Er-doped ZBLAN fiber which is the same type with the gain fiber in the seed laser. The coupling efficiency of light into the fiber amplifier was measured to be ~40% (~0.27 nJ in-fiber pulse energy), and the gain fiber length in the amplifier was 5.5 m, as illustrated in Fig. 1(a). As we gradually increased the counter-propagating pump light from 0 W to 15 W, we recorded the average power, the optical spectrum and autocorrelator traces at the amplifier output. Some typical results are plotted as blue curves in Fig. 2(a) and 2(b). For each case, the entire energy of the pulse package ($E_p$) can be calculated using the measured laser average power. It can be seen from Fig. 2(a) that as $E_p$ increased from ~10 nJ to ~35 nJ, the bandwidth of the output spectrum increased gradually. When $E_p$ is higher than 25.3 nJ, obvious Raman-induced frequency red shift of the pulse is observed. In the temporal domain (see Fig. 2(b)), when $E_p$ increase gradually from 9.7 nJ to 16.5 nJ, a sharp decrease of the pulse width has been observed. After that, strong oscillation of the pulse width has been observed, accompanying by the appearance of a sidelobe pulse with a gradually increasing intensity, see Fig, 2(b).

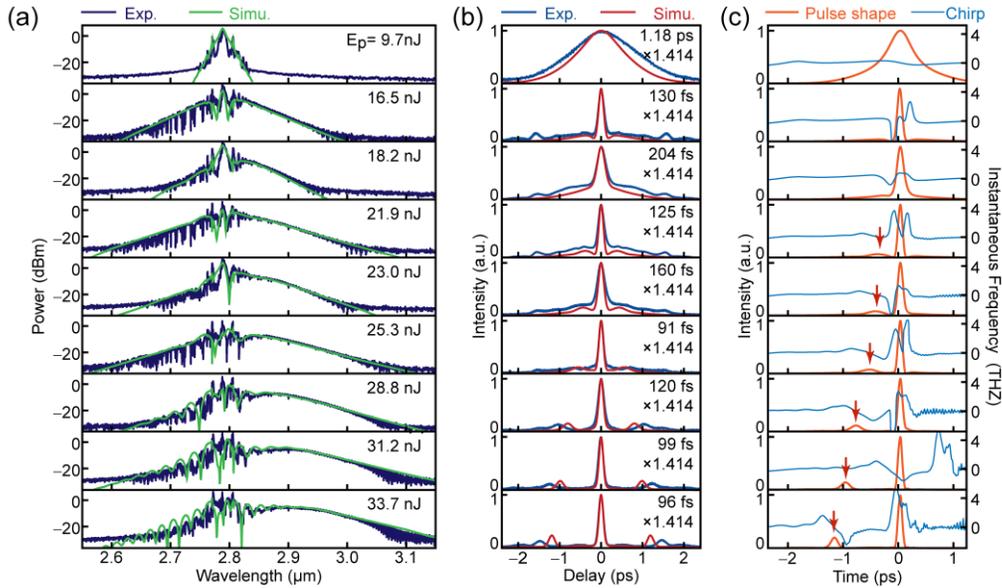

**Fig. 2** Optical spectrum and duration of the laser pulses from the amplifier. (a) Measured (dark blue) and simulated (green) optical spectra at different pulse energy. (b) Measured (blue) and simulated (red) autocorrelation traces corresponding to the spectra shown in (a).( c) Simulated pulse shapes (orange) and instantaneous frequency (light blue).

## 3 Simulations and analysis

In order to better understand the nonlinear amplification process happened in the strong anomalous gain fiber. We performed numerical simulations using the nonlinear Schrodinger equation (NLSE).[28,33] While the conventional split-step Fourier method [28] was used in the simulations, some fiber parameters were summarized in the Section 1 of the Supplementary material. Through using the pulse parameters in the experiments, we can properly retrieve the nonlinear propagation process in the gain fiber, as illustrated in Fig. 2. It can be seen that the numerical results exhibit striking agreement with the experimental ones. Some detailed numerical results are illustrated in Fig. 3, and the related experimental measurements are plotted in Fig. 3(a) for comparison.

Based on the adiabatic soliton theory, we can easily calculate the full-width-half-maximum (FWHM) temporal width ($t_{FWHM}$) of the amplified pulse at different $E_p$ using the equation of $E_p t_{FWHM} = 3.52|\beta_2|/\gamma$, where $\gamma = 0.26$ km$^{-1}$W$^{-1}$ is the Kerr nonlinearity coefficient of the gain fiber. The calculated results are plotted as the red dashed line in Fig. 3(a), and large deviation of the adiabatic soliton prediction from the experimental results is observed, which can be understood through comparing the gain and nonlinear lengths of the system (see Equation 1). At the case of 30 nJ output pulse energy, the gain length can be estimated to be ~1 m, while the nonlinear length in the gain fiber, for the input pulse (0.27 nJ, 998 fs), is ~17 m. The criterion of adiabatic soliton amplification ($L_g \gg L_n$) is far from satisfied.

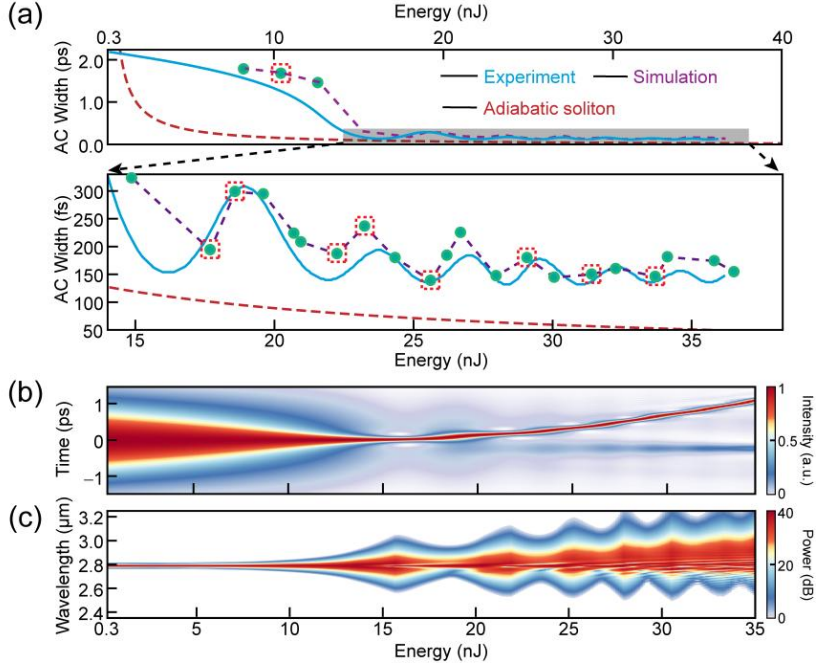

**Fig. 3.** Performance of laser amplifier as pulse energies increase. (a) Comparison of the amplifier output autocorrelation width between experiment (green point and purple dash) and simulation (blue solid line). The calculated fundamental soliton duration (red dash line) is significantly shorter than both experimental and simulation results. (b) and (c) Temporal and spectral evolution of the pulse with increasing pulse energy.

To illustrate the nonlinear amplification dynamics in the gain fiber with strong anomalous dispersion, we performed numerical simulations of in-fiber pulse propagation at the case of ~30 nJ output pulse energy (corresponding to ~900 mW output average power). The simulation results are illustrated in Fig. 4. While the pulse energy variation over the 5.5-m-long gain fiber is illustrated in Fig. 4(a), the simulated evolutions of pulse temporal profile and FWHM pulse width are illustrated in Fig. 4(b) and 4(c), respectively. In the simulation, we used a quality factor as $\tau/\tau_{TL}$ to evaluate the pulse chirp and pedestal performance, which can be expressed as the ratio of the actual pulse width ($\tau$) to its Fourier transform-limit (TL) width ($\tau_{TL}$), see Fig. 4(d). The TL pulse width can be calculated directly using the simulated optical spectrum of the pulse (see Fig. 4(e)), when assuming the pulse chirp is zero.

*3.1 Stage 1: Balance between linear and nonlinear chirp*

It can be found from Fig. 4 that the nonlinear amplification process in the gain fiber can be divided into three distinct stages, each of which exhibits unique features in both temporal and spectral domains. The first stage is from the beginning to ~3.4 m of the gain fiber, in which the pulse energy increases from 0.27 nJ to 5.3 nJ. In this stage, the basic mechanism of pulse propagation is the balance between the accumulated linear chirp (due to the fiber dispersion) and the accumulated nonlinear chirp (due to the self-phase modulation effect). It can be seen from Fig. 4(b) that at the first ~2.2 m of the gain fiber, the nonlinear effect is rather weak due to the relatively-low pulse energy, meanwhile the pulse temporal width increases slightly due to the accumulated negative chirp in the anomalous dispersion fiber, giving a gradually increasing quality factor ($\tau/\tau_{TL}$), see Fig. 4(d). As the pulse energy increases in the gain fiber, the strength of the self-phase modulation (SPM) effect increases, which can lead to a positive pulse chirp, compensating the negative chirp due to fiber dispersion. Such compensation can result in a decrease of the quality factor to its local minimum at a fiber position of ~3.4 m, see Fig. 4(d). At this local-minimum position, the amplified pulse has a pulse energy of ~5.3 nJ and a pulse width of ~770 fs, giving a soliton order of ~2.

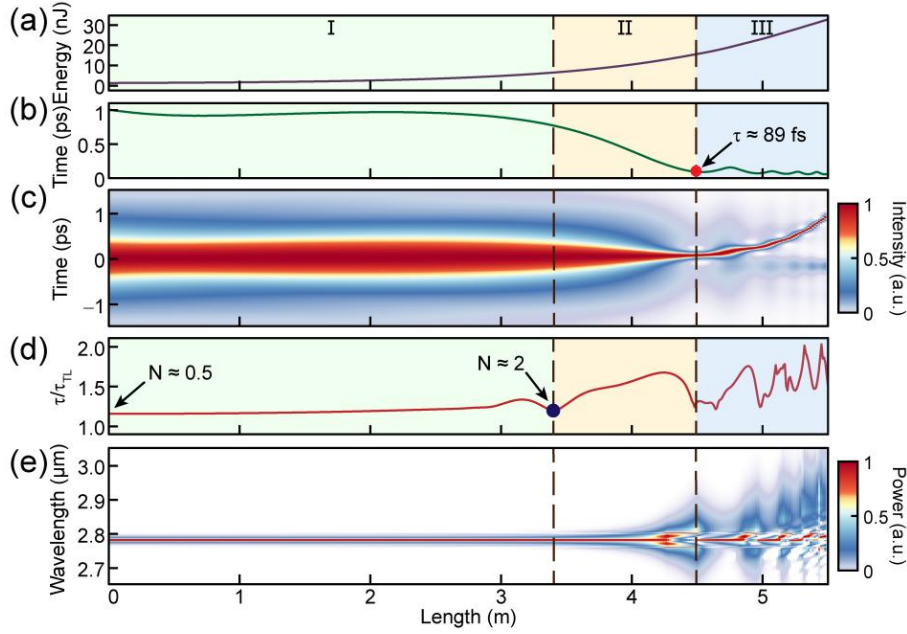

**Fig. 4.** Three-stage nonlinear amplification process. (a) and (b) The evolution trend of pulse energy and duration. (c) Simulated temporal evolution of the laser pulse in the ZBLAN fiber. (d) $\tau/\tau_{TL}$ variation (see text for details. (e) Simulated spectral evolution.

*3.2 Stage 2: High-order-soliton-like pulse compression*

From ~3.4 m to ~4.5 m in the gain fiber, the pulse experiences a sharp temporal compression over amplification. The compression process can be described using the well-known theory of high-order soliton compression. The strong SPM effect leads to a broadening of the pulse spectrum. While at the leading edge of the pulse the SPM-induced index modulation moves the optical components to the lower-frequency side, the optical components at the pulse trailing moves to the higher-frequency side due to the same effect. In anomalous dispersion fiber, the lower-frequency component has a lower group velocity than the higher-frequency one, therefore optical components on both sides of the pulse edges moves toward the pulse temporal center, resulting in fast temporal compression of the high-order soliton. It can be found in Fig. 4(c) that this high-order-soliton-like pulse compression leads to a sharp decrease of the pulse width from 0.8 ps to 89 fs over a short fiber length of ~1 m.

To further verify the fact that this sharp decrease of pulse width at the second stage results from a high-order-soliton-like compression process, we performed some additional simulations. First, we started the pulse propagation simulation from the beginning point ($L = 3.4$ m) of the second stage, and switched off the gain term as well as all the high-order ($\geq$3-order) dispersion and Raman nonlinearity terms, leaving merely the second-order dispersion and Kerr nonlinearity in the model. At this condition, the simplified nonlinear Schrodinger equation supports canonical solutions of high-order solitons which exhibit periodic spectral and temporal breathing [23]. As illustrated in Fig. 5(a), the feature of periodicity for the high-order soliton, in both temporal and spectral domains, has been clearly observed when we extend the fiber length in the simulation to 14 m.

When the gain term is switched on (keeping high-order dispersion and Raman nonlinearity off), we simulated the pulse propagation process and the results are illustrated in Fig. 5(b). It can be found that with the gain term added, the periodicity feature of the pulse evolution remains, exhibiting however an obvious increasing of the breathing period. This breathing period increase can be easily understood as: as the pulse energy increases due to the fiber gain, the soliton order ($N$) gradually increases as $N^2 \sim (\gamma E_P \tau)/|\beta_2|$, leading to the decreasing of both the nonlinear length and the breathing period, see Fig. 5(b). After switching on all the high-order dispersion and Raman nonlinearity terms in the simulation, we found that both the temporal and spectral evolutions of the pulse changes dramatically, especially after the first compression point at ~4.5 m of the gain fiber, see Fig. 5(c).

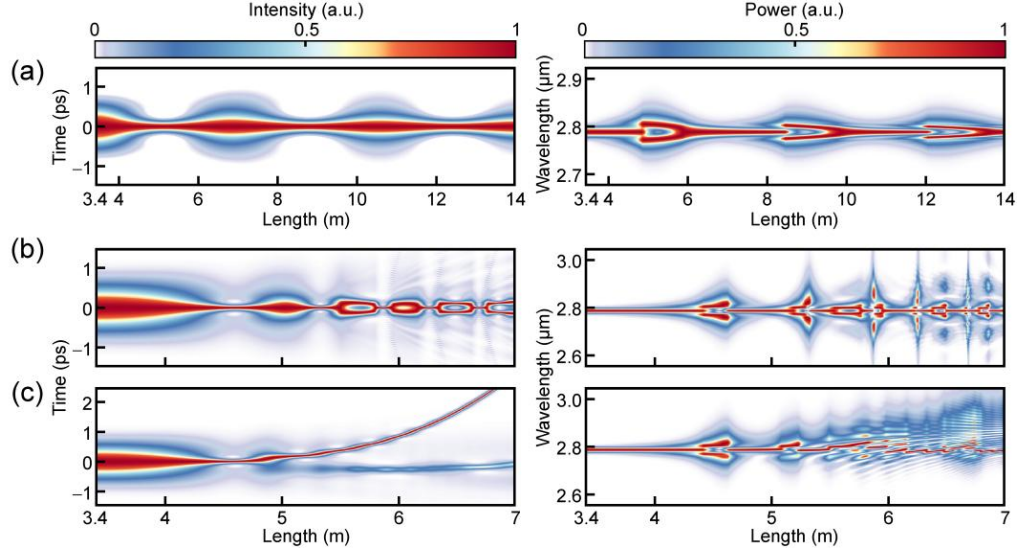

**Fig. 5.** Higher-order soliton dynamics under different conditions. (a) Without gain or high-order effects.( b) With gain only. (c) With both gain and high-order effects.

### 3.3 Stage 3: Pulse splitting due to high-order effects

The third stage of pulse propagation results from the high-order-soliton-like compression in the second stage, which gives rise to largely-enhanced pulse peak power as well as largely-broadened spectral coverage, leading to therefore strong high-order effects. High-order (≥3-order) dispersion and Raman nonlinearity breaks both the spectral and temporal symmetries of the nonlinear system, resulting in the splitting of the high-order soliton, [34] see Section 2 of the Supplementary material file. As illustrated in Fig. 4(c), 4(e) and 5(c), the high-order soliton splitting causes the generation of a Raman self-frequency shift soliton with some oscillations on both its temporal width and spectral bandwidth, meanwhile a residual pulse leaves at the wavelength of the seed pulse as the pedestal, whose amplitude is amplified over propagation. The higher-order-effect-induced soliton splitting also causes obvious degradation of the pulse quality, giving rise to an increase of the $\tau/\tau_{TL}$ factor in the third stage, see Fig. 4(d). It should be noted that even though the Raman self-frequency shift soliton, generated from high-order-soliton splitting, becomes a fundamental soliton after few-meter-length propagation in the third stage (see Fig, 5(c)), such a soliton-like propagation after temporal and spectral oscillations is still quite different from the theory of adiabatic soliton amplification. This is because the Raman effect moves the central wavelength of the pulse away from the fiber gain spectrum (see Fig. 6(c)), and the increases of pulse energy near the end section of the third stage comes mainly from the amplification of the pulse pedestal.

The effect of Raman soliton self-frequency shift and the amplification of pulse pedestal together, clamp the pulse peak power achievable in the gain fiber. As illustrated in Fig. 6(a), we plot the ratio of the main pulse energy to the total energy in the entire pulse packet and the peak power of the main pulse at different gain level, when the gain fiber length remains to be 5.5 m. We found that in the simulation, when we gradually increase the total output pulse energy from 10 nJ to 50 nJ, the energy ratio of the main pulse decreases from ~1 to <0.6, while the pulse peak power is clamped at ~250 kW.

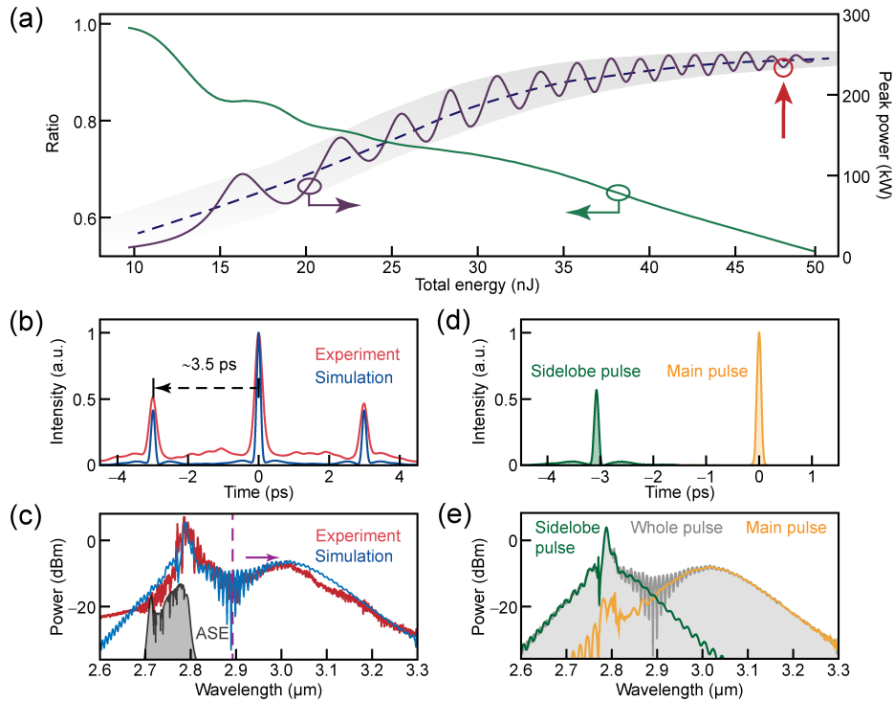

**Fig.6.** Pulse splitting induced by Raman soliton self-frequency shift. (a) Peak power (purple) and energy proportion (green) of the main pulse. (b) Measured (red) and simulated (blue) autocorrelation traces. (c) Measured (red),

simulated (blue) spectra of output pulse and measured amplified spontaneous emission (ASE, gray) at high power pump. (d) and (e) The simulated temporal and spectral characteristics of the main pulse (yellow) and sidelobe pulse (green).

The increase of the pulse pedestal, observed in the simulation, has been verified in the experiment, see Fig. 2(b) and 6(b). At an output pulse energy of 50 nJ, we observed experimentally a sidelobe pulse (pedestal) with nearly half intensity of the main pulse (see Fig. 6(b)). The experimental results can be perfectly retrieved in the simulations, see Fig. 6(b) and 6(c). While the sidelobe pulse can be considered as the pulse pedestal with a central wavelength of ~2.8 μm, the main pulse is the Raman self-frequency shift soliton which has a central wavelength of ~3 μm (outside the fiber gain spectrum), see Fig. 6(d) and 6(e).

## 4. Universality of the three-stage dynamics

This three-stage model indicates that for the nonlinear pulse amplification process in strong anomalous gain fiber, there is an optimized output position which locates at the end of the second stage (with the minimum pulse width due to high-order-soliton-like compression). At this optimized position, relatively-clear compressed pulse can be obtained, while detrimental high-order effects, that can lead to soliton splitting and Raman self-frequency shift, are largely suppressed. In order to verify the universality of the three-stage model and to provide some guidelines for designing the fiber amplifier, we performed some simulations when maintaining the input pulse parameters (1 ps, 0.27 nJ, $N_{start}$~0.5) launched into the fiber amplifier. As we gradually increase the gain coefficient (decrease the gain length), it can be found that the optimized output position moves toward the input port of the gain fiber and the output pulse energy increases, see Fig. 7(a). These phenomena can be understood as: the increase of gain coefficient results in a faster increase of the pulse energy, and therefore the balance between the linear (due to dispersion) and nonlinear pulse chirp (due to SPM effect) can be realized over a shorter gain fiber length. In the second stage of high-order-soliton-like pulse compression, a higher pulse energy leads to a higher soliton order, giving rise to a faster pulse compression process, as illustrated in Fig. 7(a). The quality factor ($\tau/\tau_{TL}$) evolutions over gain fiber length, for all these three cases, exhibit quite similar manners, highlighting further the universality of the three-stage model.

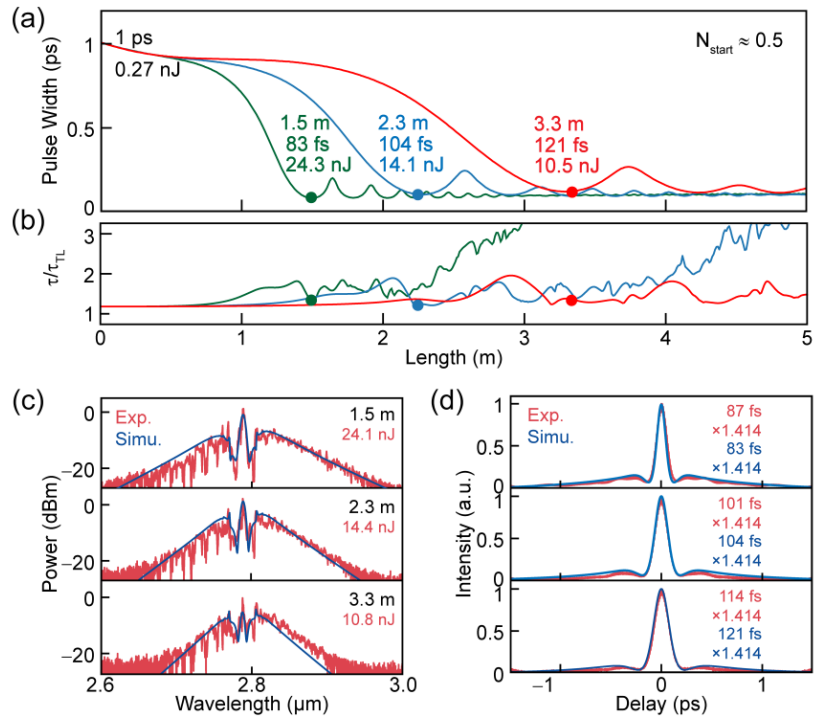

**Fig. 7.** Three-stage pulse evolution model under different gain coefficients. (a) and (b) Evolution of pulse duration and quality factor $\tau/\tau_{TL}$. (c) and (d) Measured (red) and simulated (blue) spectra and autocorrelation traces at the first compression points.

The guideline for the amplifier design, based on the three-stage-model understanding, can be directly used for optimizing the output performance of the system. As illustrated in Fig. 7(c) and 7(d), as we gradually increased the pump power of the amplifier and shortened accordingly the gain fiber length in the experiment, relative-clear ultrafast pulses with compressed pulse widths and low pulse pedestals can be obtained directly at the output port of the gain fiber. The experimentally measured optical spectra and autocorrelation traces of the output pulses, for all the three output pulse energies, exhibit striking agreement with the theoretical predictions. For these three cases, the energy portions within the main pulses are estimated to be 81%, 85% and 87% respectively, and the quality factors ($\tau/\tau_{TL}$) of the output pulses are all ~1.1. See Section 3 of the Supplementary material file for more detailed theoretical and experimental results.

To further verify the universality of the model, we maintained the gain coefficient (gain length) of the fiber amplifier and varied the input pulse parameters in the simulation. As illustrated in Fig. 8(a) and 8(b), when the input pulse energy and pulse width are varied to be 0.47 nJ and 560 fs (the soliton order remains to be ~0.5), the optimized gain fiber length is shortened to be 1.5 m, mainly due to the fact that the higher pulse energy and shorter pulse width strongly enhance the nonlinearity in the gain fiber, shortening the gain fiber lengths required for both the first- and the second-stage amplification processes. Even though large fiber-length scaling (from 1.5 m to 3.3 m) has been observed for different input power parameters (see Fig. 8(a)), the very similar manners of evolutions (three-stage dynamics), for both the pulse width (Fig. 8(a)) and the pulse quality factor (Fig. 8(b)), exhibit clearly three-stage dynamics with different length scales. In practice, we performed the amplification experiments using 560 fs solitonic pulse with a pulse energy of ~0.47 nJ as the seed light, and adjusted both the pump power and the gain fiber length to optimize the output performance of the amplifier. The experimental results exhibit perfect agreement with the theoretical prediction based on the three-stage model, see Fig. 8(c) and 8(d) and Section 3 of the Supplementary material file. The autocorrelation measurements (see Fig. 8(d)) highlight the high-quality ultrafast pulse generation at the output port of the amplifier with low pulse pedestals, giving rise to high energy portions (~89% and ~87%) at the main pulses.

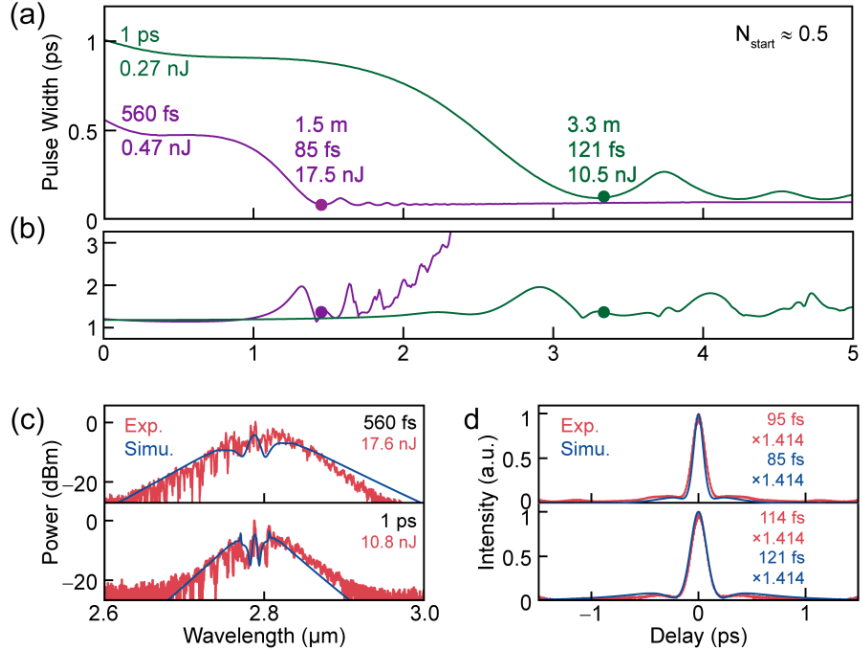

**Fig. 8.** Three-stage pulse evolution model under different input pulse parameters. (a) and (b) Evolution of pulse duration and quality factor $\tau/\tau_{TL}$. (c) and (d) Measured (red) and simulated (blue) spectra and autocorrelation traces at the first compression points.

## 5. Discussion and conclusion

It can be found that in the third stage of this three-stage dynamics, the phenomenon of soliton splitting results in the generation of an ultrafast main pulse, which can be regarded as a fundamental soliton whose central frequency continuously shifts to longer wavelengths due to Raman effects. We would like to mention here that the generation of this output soliton can hardly be understood using the theory of adiabatic soliton amplification, and the output pulse parameters exhibit significant deviations from the results predicted by the adiabatic theory, see Fig. 3(a). Moreover, we notice from the three-stage dynamics that the generation of this fundamental soliton, the appearance of large pulse pedestal and the phenomenon of Raman-induce self-frequency shift, all of them are the consequences of high-order-soliton splitting, which is inevitable in the amplification process if a fundamental-soliton-like pulse is expected at the output port of the gain fiber. Contrarily, in order to obtain a high-performance ultrafast output pulse with compressed temporal width and low pedestal, the three-stage model points out an optimized gain fiber length, at which the second stage of high-order-soliton-like compression just finishes and excessive nonlinearity, high-order dispersion effects and high-order-soliton splitting can be efficiently avoided.

The three-stage model unveils the concept that the first compression point of the high-order-soliton-like compression stage is the key for the nonlinear amplification process in anomalous dispersion gain fiber to obtain high-performance output pulse. The guidelines for fiber amplifier design based on this model, are quite different (sometimes being opposite) with the guidelines from the adiabatic soliton amplification theory. For example, in order the obtain a higher output pulse energy, the three-stage model suggests to shorten the gain fiber length and at the same time to increase the fiber gain coefficient (see Fig. 7(a)), while based on the theory of adiabatic soliton amplification, a longer gain fiber length would be necessary for a higher output pulse energy so as to ensure the adiabaticity of the pulse propagation.

Moreover, the three-stage model can give simple and insightful interpretation about the pre-chirp engineering method for the nonlinear amplification systems in anomalous dispersion gain fiber, which has been frequently used in previous experiments.[9,10] However, the underlying mechanism of this pre-chirp method is inexplicit so far. In the three-stage model, the method of pre-chirp engineering is to control the length of the first stage of the system (to adjust the balance between linear and nonlinear pulse chirp) through introducing an additional pulse chirp to the seed pulse. This chirp control, in one side, is an efficient means of adjusting the optimized position of the second stage of high-order-soliton-like compression, in the other side it can also vary the pulse energies both at the beginning of second-stage pulse compression and at the optimized gain fiber output.

In summary, we demonstrate that the nonlinear amplification process in a section of anomalous dispersion gain fiber, can be divided into three distinct stages: the balance between linear and nonlinear chirp, high-order-soliton-like compression, and pulse splitting due to high-order effects. We unveil both theoretically and experimentally that this nonlinear amplification process has an optimized gain fiber length, at which compressed ultrafast pulses can be obtained with low pulse pedestals and relatively-weak Raman frequency shifts. While a longer gain fiber length, leading to excessive nonlinearity, can hardly enhance the peak power of the amplified pulse, some new guidelines for fiber amplifier design have been pointed out according to this three-stage model. Our results provide some physical insights on the nonlinear amplification process in high-gain fiber with anomalous dispersion, and the demonstrated three-stage model could be useful for designing high-performance nonlinear fiber amplifiers at near- and mid-infrared wavelengths.

**Funding Information.**

The authors acknowledge support from Strategic Priority Research Program of the Chinese Academy of Science (XDB0650000); Shanghai Science and Technology Innovation Action Plan (21ZR1482700); Shanghai Science and Technology Plan Project Funding (23JC1410100); National High-level Talent Youth Project; Fuyang High-level Talent Group Project.

**Disclosures.**

The authors declare no conflicts of interest.

**Data availability.**

Data underlying the results presented in this paper are not publicly available at this time but may be obtained from the authors upon reasonable request.

**Weiyi Sun** is a PhD student at ShanghaiTech University, China. He received his BS degree from Zhejiang University. His research focuses on high-power ultrafast mid-IR fiber laser and nonlinear dynamics in fiber amplification.

**Jiapeng Huang** is a professor at Shanghai Institute of Optics and Fine Mechanics (SIOM), Chinese Academy of Sciences (CAS), Shanghai, China. He earned his doctor degree in 2020 from Universität Erlangen-Nürnberg and the Max-Planck Institute for the Science of Light, Germany. His research focuses on fabrication of soft-glass photonic crystal fibers, high-power ultrafast mid-infrared fiber laser and mid-infrared fiber sensors.

**Liming Chen** is a PhD student at SIOM, CAS, China. He received his master degree from ShanghaiTech University, China. His research focuses on high-power mid-infrared fiber components and fiber lasers.

**Zhuozhao Luo** is a post-doctoral researcher at SIOM, CAS, China. She received her MS and PhD degree from Wuhan University of Technology. Her research focuses on design and fabrication of novel hollow-core photonic crystal fibers.

**Wei Lin** is a PhD student at Hangzhou Institute for Advanced Study, University of CAS ,China. He received his BS degree from Ningbo University. His research focuses on ultrafast mid-infrared fiber laser.

**Zeqing Li** is a PhD student at SIOM, CAS, China. She received her BS degree from Tianjin University. Her research focuses on nonlinear dynamics in ultrafast mid-infrared fiber laser.

**Cong Jiang** is a PhD student at SIOM, CAS, China. He received his MS degree from Jinan University. His research focuses on optical frequency comb in mid-infrared region.

**Zhiyuan Huang** is a professor at SIOM, CAS, China. He received his PhD degree from SIOM. His research focuses on ultrafast laser technology, few-cycle pulse compression, and broadband ultrafast ultra-violet light source.

**Xin Jiang** is a professor at Hangzhou Institute of Optics and Fine Mechanics, China. He received his PhD degree from University of Leeds, UK. His research focus on the fabrication of novel photonic crystal fibers and related applications, including supercontinuum light sources, photochemistry, quantum optics, fiber optic sensing, optical tweezers.

**Pengfei Wang** is a professor from Northeast Normal University, China. He received his PhD degree from Technological University Dublin, Ireland. His research focus on mid-infrared fiber laser and mid-infrared optical glasses.

**Yuxin Leng** is a professor at SIOM, CAS, Shanghai, China, with interests in ultra-intense lasers and physics. He received his bachelor's and doctoral degrees from Wuhan University and SIOM in 1997 and 2002, respectively. He is the director of State Key Laboratory of High Field Laser Physics, SIOM, CAS and vice-director of SIOM, CAS.

**Meng Pang** is a professor at SIOM, CAS, China. He received his MS degree from Tsinghua University and PhD degree from Hong Kong Polytechnic University. His research focuses on nonlinear fiber optics, ultrafast fiber laser and photonic crystal fibers.